\documentclass[sigconf]{acmart}

\def\BibTeX{{\rm B\kern-.05em{\sc i\kern-.025em b}\kern-.08emT\kern-.1667em\lower.7ex\hbox{E}\kern-.125emX}}
    
\copyrightyear{}
\acmYear{}
\setcopyright{none}
\acmConference[KDD '19]{SIGKDD '19}{August 04--08, 2019}{Anchorage, Alaska}
\acmPrice{}
\acmDOI{}
\acmISBN{}

\usepackage{graphicx} 
\usepackage{subfigure} 

\usepackage{natbib}

\usepackage{algorithm}
\usepackage{algpseudocode}
\usepackage{hyperref}
\usepackage{amsthm,amsmath,amssymb,amsfonts,bbm,graphics,a4wide,rotating}
\usepackage{verbatimbox,enumitem}

\definecolor{darkred}{rgb}{.8,.0,.035}

\def\ones{\mathbbm{1}}

\def \reals{\mathbb{R}}
\def \expect{\mathbb{E}}

\newcommand{\vect}{\boldsymbol} 

\def\vz{\vect{z}}
\def\vq{\vect{q}}
\def\vx{\vect{x}}

\def\ve{\vect{\mu}}

\begin{document}
\title[AI-Augmented Lesion Detection]{An AI-Augmented Lesion Detection Framework For Liver Metastases With Model Interpretability}

%

\author{Xin J. Hunt}
\affiliation{%
  \institution{SAS Institute Inc.}
}
\email{xin.hunt@sas.com}

\author{Ralph Abbey}
\affiliation{%
  \institution{SAS Institute Inc.}
}
\email{ralph.abbey@sas.com}

\author{Ricky Tharrington}
\affiliation{%
  \institution{SAS Institute Inc.}
}
\email{ricky.tharrington@sas.com}

\author{Joost Huiskens}
\affiliation{%
  \institution{SAS Institute B.V.}
}
\email{joost.huiskens@sas.com}

\author{Nina Wesdorp, MD}
\affiliation{%
  \institution{Amsterdam University Medical Centers}
}
\email{n.wesdorp@amsterdamumc.nl}

%
\renewcommand{\shortauthors}{Hunt, Abbey, Tharrington, Huiskens, and Wesdorp}

%
\begin{abstract}

Colorectal cancer (CRC) is the third most common cancer and the second leading cause of cancer-related deaths worldwide. Most CRC deaths are the result of progression of metastases. The assessment of metastases is done using the RECIST criterion, which is time consuming and subjective, as clinicians need to manually measure anatomical tumor sizes. AI has many successes in image object detection, but often suffers because the models used are not interpretable, leading to issues in trust and implementation in the clinical setting. We propose a framework for an AI-augmented system in which an interactive AI system assists clinicians in the metastasis assessment. We include model interpretability to give explanations of the reasoning of the underlying models.
\end{abstract}

%
%

%

%

%

%
\maketitle

\section{Introduction}
\label{sec:introduction}
Colorectal cancer (CRC) is the third most common cancer and the second leading cause of cancer-related deaths worldwide~\cite{bray2018global}. Most cancer deaths are the result of progression of metastases. Approximately 50\% of CRC patients will develop metastases to the liver (CRLM)~\cite{abdalla2006improving, donadon2007new}. Patients with liver-only colorectal metastases can be treated with curative intent. Complete surgical resection of CRLM is considered the only method with a chance to cure these patients~\cite{noren2016selection, de2016management, angelsen2017population}. Only 20\% of the patients with CRLM present with resectable CRLM~\cite{wicherts2007bringing, d2009treatment, poston2008urgent, adam2004rescue}. Initially-unresectable liver metastases can become resectable after downsizing the lesions via systemic therapy. However, there is no consensus regarding the optimal systemic therapy regime. The effect of systemic treatment varies between patients, some have total response and others show progression of disease~\cite{adam2004rescue, guideline2014}. Moreover, systemic therapy has a lot of side effects due to its cytotoxicity~\cite{meyerhardt2005systemic}. 

In clinical oncology, the selection and monitoring of treatment is crucial for effective cancer treatment and for the evaluation of new drug therapies. Accordingly, assessment of patient response to treatment is a crucial feature in the clinical evaluation of systemic therapy. The widely accepted and applied criterion for such assessment is the Response Evaluation Criteria In Solid Tumors (RECIST), which aims to measure the objective change of anatomical tumor size. The RECIST assessment is performed by measuring changes in one-dimensional (1-D) diameter in two target lesions before and after therapy~\cite{eisenhauer2009new}. Though RECIST is a clinical standard worldwide, it is highly limited. Currently, it is not possible to predict clinical outcome based on tumor response assessment (RECIST) and patient characteristics in individual patients. A meta-analysis revealed that inter-observer variability in RECIST measurement may exceed the 20\% cut-off for progression, resulting in potential misclassification of diagnosis (stable disease or progression)~\cite{yoon2016observer}. A further problem of RECIST is the subjectivity and variability in selecting target lesions. 

%

\subsection{Assessment Automation}
One of the goals of this project is to more accurately and efficiently assess tumor response. Automated medical image processing allows more objective analysis of clinically relevant imaging features. Machine learning methods like object detection models trained on clinical data can be used to detect tumor lesions and automate systemic therapy response assessment. However, fully automated systems directly based on object detection models are not ideal under current circumstances, for reasons including
\begin{itemize}[leftmargin=*]
\item Accuracy concerns: Modern general purpose object detection methods can only achieve 30\% to roughly 75\% mean average precision (mAP), depending on the dataset they are tested on\cite{huang2017speed}. Specialized models trained specifically for lesion detection may achieve better accuracy, but will still make mistakes. 
Incorporating professional knowledge and feedback from clinicians can significantly improve detection accuracy.
\item Lack of interpretability: Most of the high-performance object detection models are based on deep learning, which are often considered black-box models. However, it is important to communicate the reason behind decisions to the physician prescribing treatment plans and to the patient. A system without the ability to explain its decisions is not desirable in the clinical setting.
\end{itemize}
Thus, in this paper we propose an interactive system, in which we use machine learning object detection models along with model interpretability and natural language generation to augment the efficacy and accuracy of clinicians in lesion detection and assessment. Model interpretability with natural language generation acts to provide trust and understanding of the underlying machine learning object detection models.

\section{Related Work}
\subsection{Clinical studies}
Artificial intelligence is quickly moving forward in many fields, including in medicine. Deep learning techniques have delivered impressive results in image recognition. Radiology and pathology are medical specialties that create and evaluate lots of medical images. Multiple studies in these specialties have used artificial intelligence to mimic or augment human capabilities. 

Masood et al.~\cite{masood2018computer} propose a computer-assisted decision support system with the potential to help radiologists in improving detection and diagnosis decision for pulmonary cancer stage classification. Wang et al.~\cite{wang2016deep} show that using a deep learning model for automated detection of metastatic breast cancer in lymph node biopsies reduces the error rate of pathologists from over three percent to less than one percent. Hamm et al.~\cite{hamm2019deep} use a convolutional neural network for fast liver tumor diagnosis on multi-phasic MRI images.



While these initial results are positive, substantial translation or implementation of these technologies into clinical use has not yet transpired~\cite{he2019practical}. Beyond building AI algorithms, applying them in daily clinical practice is complex~\cite{dreyer2017machines}. Key challenges for the implementation include data sharing, patient safety, data standardization, integration into complex clinical workflows, compliance with regulation, and transparency~\cite{he2019practical}. Transparency of complex AI algorithms is of great importance for clinicians. If a medical doctor cannot understand the outcome of an algorithm, then the doctor will be unable to explain the outcome to a patient. Technologies that help explain complex AI algorithms have an important role in acceptance of AI by the medical community.  

\subsection{Model interpretability}
In our proposed framework there is a need for an applicable model
interpretability method. The majority of methods for model
interpretability come in three forms:
inherently-interpretable models \cite{CORELS, lou2013accurate,
CynthiaScorecards}, methods for interpreting existing models
\cite{montavon2018methods, samek2017explainable}), and post-hoc
investigations \cite{SHAP, LIME, ribeiro2018anchors, sampleSHAP}.
In the current framework, we use a segmentation convolutional neural network model for lesion detection in CT images for the CRLM patients. To give explanations of this model, we use Shapley values, a post-hoc, model-agnostic interpretability method. The model-agnostic nature of the Shapley values gives us the flexibility to substitute better performing models in future research.

The Shapley Values were originally introduced in game theory as a way to
determine the individual contributions of players in a collaborative
game. In model interpretability the Shapley values are used to measure
the individual contributions of the input variable values of a single
observation to a model's prediction. Recent work in this area includes
\cite{SHAP, sampleSHAP}. Image models use as inputs a series
of pixel values. Assigning Shapley values to these pixels creates a gradient over
an image indicating regions of an image that lead to or detract from lesion detection
probabilities. This gradient is very easy to understand, which makes Shapley values
a natural method for the clinician-collaboration framework.

\section{Proposed Framework}
\label{sec:framework}
\begin{figure}[ht!]
\centering
\includegraphics[width=.35\textwidth]{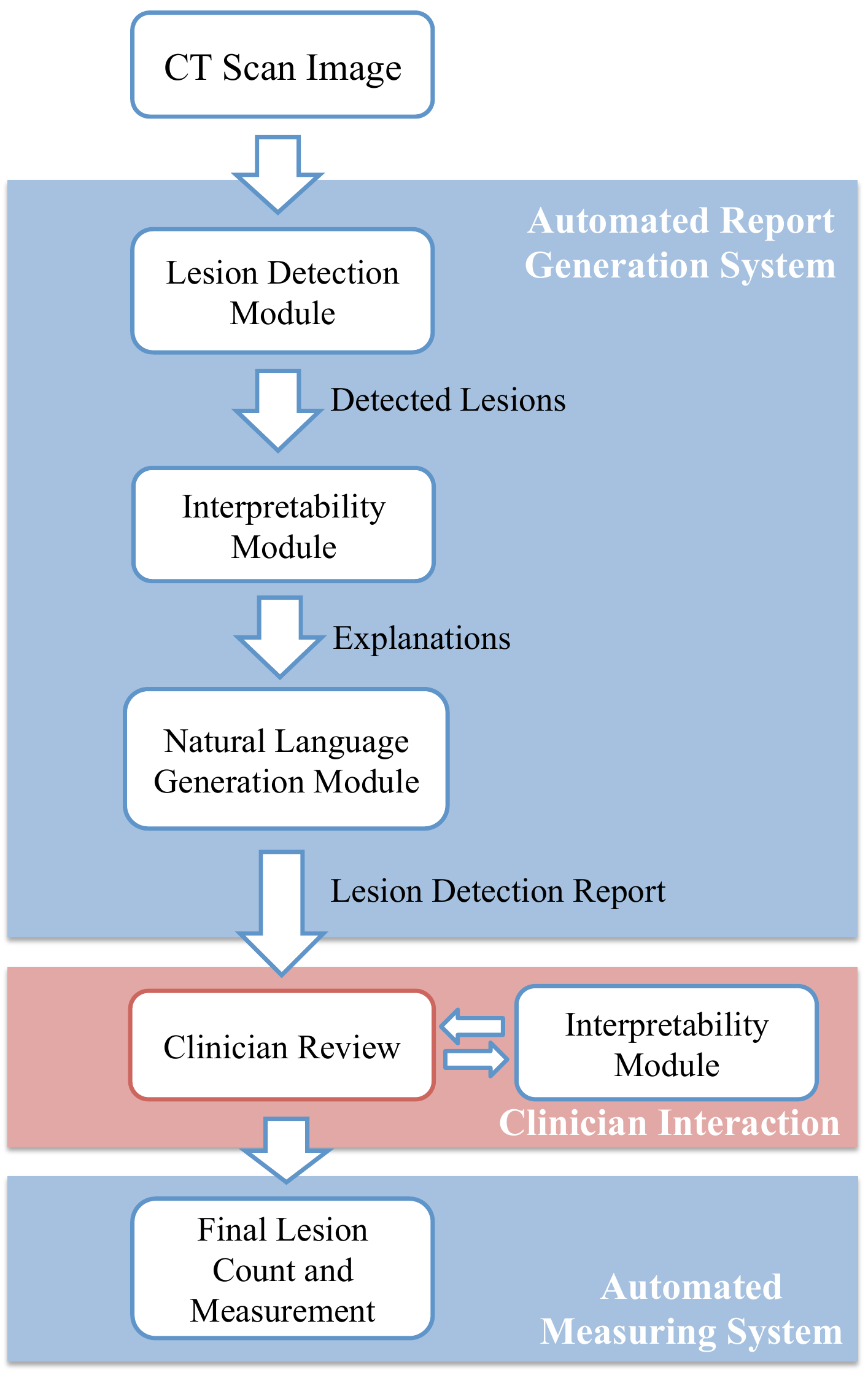}
\caption{Flow chart of the proposed system}
\label{fig:flowchart}
\end{figure}

In this section we propose a system with clinician interaction for high accuracy lesion detection and measurement. The proposed system is summarized in Figure~\ref{fig:flowchart}. 

The images from a CT scan are sent to an automated report generation system. The report generation system's purpose is to create an interactive and coherent report highlighting possible lesions. This report will allow a clinician to quickly confirm or overwrite the detections made by the automated system. The report generation system consists of three modules: a lesion detection module, an interpretability module, and a natural language generation module. The lesion detection module processes the scan images and labels all potential lesions. 
Note that we can use any high-accuracy object detection model as the detection module. 
The detected lesions are sent to the interpretability module to generate visual explanations that highlight important decision areas, which can help the clinician make better informed decisions. The natural language generation module then collects information from previous modules and generates a report for the clinician.

Once a report is generated, a clinician then reviews the report, confirming or rejecting each individual detection. The clinician can also add new lesion detections missed by the automated system. During the review process, the clinician can interact with the interpretability module, review explanations for detected lesions, and request explanations for new areas as indicated by the clinician. Once all detections are confirmed or rejected, the scan image is sent to the automated measuring system, where information like lesion count, location, and diameter is recorded.


\section{HyperSHAP (\textit{Hyper}parameterized \textit{SHAP}ley value estimation)}
\label{sec:algorithm}

The interpretability module in our proposed framework uses Shapley values.  We propose a novel and deterministic approximation to the Shapley values for efficient computation.

The calculation of the Shapley values for the $i^{\mbox{th}}$ variable of an instance of interest, which we call a query $\vq$, given a predictive model $f:\reals^m \rightarrow \reals$ is
\begin{equation}
\begin{aligned}
 \phi_i = 
 \sum_{\vz \in [0,1]^m, \vz_i = 0} \frac{(m - |\vz| - 1)!|\vz|!}{m!}\left( \expect_{\vz + \vect{e}_i}[(f(\vx)] - \expect_{\vz}[f(\vx)] \right),
\end{aligned}
\notag
\end{equation}
where the expectation
$
\expect_{\vz}[f(\vq)] = \frac{1}{n} \sum_{\vx \in X}f(\tau(\vq, \vx, S))
$
is computed over all observations $\vx$ in a data set $X$ with $n$ total observations. The function
$
\tau(\vq, \vx, S) = \begin{cases} q_i \quad i\in S \\
x_i \quad i \not\in S
\end{cases},
$
and the relationship between $\vz$ and $S$ is
$
z_i = \begin{cases} 1 \quad \mbox{variable  } i \in S \\
0 \quad \mbox{variable  } i \not\in S
\end{cases}.
$

$S$ represents a subset of variables used in the model training. The summation in the Shapley value computation is over all $\vz \in [0,1]^m, \vz_i = 0$, that is to say, all subsets of variables that are not the variable of interest.

It is important to note that the Shapley value computation requires all subsets of variables, of which there are $2^m$. This computation complexity makes direct computation infeasible. Instead, we rely on a deterministic approximation that uses only allowed subsets of variables as determined by a hyperparameter $\chi$, leading to the name HyperSHAP. We use a deterministic approximation to ensure stability in explanations, while still achieving high accuracy of the Shapley values.

Algorithm~\ref{alg:hyperShap} describes the full HyperSHAP computation for a value of $\chi$, using Algorithm~\ref{alg:expectation}, which computes the expectations.

%
\begin{algorithm}[h!]
\caption{Shapley Expected Values}
\begin{algorithmic}[1]
    \State {\bf input:}  training data $X^t\in\reals^{n\times m}$, query data $\vect{q}\in\reals^m$, 
    model function $f(\cdot):\reals^m \rightarrow \reals$,
    selection matrix $Z\in\{0,1\}^{c\times m}$
    \State {\bf initialize:} $\ve = {\bf 0}_{c\times 1}$ be an all-zero vector
    \For{$j=1\ldots,c$}
        \State {\bf initialize:} $X^{\vz} = [\ ]$
        \State Let $\vect{z} = Z_{[j,:]}$ be the $j^{\mbox{th}}$ row of $Z$
        \For {$i=1\dots, n$}
            \State Compute $\vx^{\vz} = \tau(\vq, X^t_{[i,:]}, \vz)$ using the $i^{\mbox{th}}$ row of $X^t$
            \State $X^{\vz} = \left[\begin{array}{c}X^{\vz} \\ \vx^{\vz} \end{array}\right]$
        \EndFor
        \State Compute $\vect{y}\in\reals^{n}$, where ${y}_t = f(X_{[t,:]}^{\vz}), t=1,\ldots,n$
        \State $\mu_j = \frac{1}{n}\ones_{n\times 1}^{\top}\vect{y}$
    \EndFor
    \State {\bf output:} $\ve$
\end{algorithmic}
\label{alg:expectation}
\end{algorithm}

\begin{algorithm}[h!]
\caption{HyperSHAP}
\begin{algorithmic}[1]
    \State {\bf input:}  training data $X^t\in\reals^{n\times m}$, query data $\vect{q}\in\reals^m$, 
    model function $f(\cdot):\reals^m \rightarrow \reals$, approximation depth $\chi$
    \State {\bf initialize:} $Z = [\ ]$
    \For{$k=0,\ldots,\chi,m-\chi,\ldots,m$}
        \State Let $Z^k\in\{0,1\}^{\binom{m}{k}\times m}$ be the selection matrix whose rows form the set $\left\{\vz\in\{0,1\}^m:|\vz| = k\right\}$
        \State $Z = \left[\begin{array}{c}Z \\ Z^k\end{array}\right]$
    \EndFor
    \State Use Algorithm~\ref{alg:expectation} to compute $\ve$ with $Z$
    \State Let $c$ be the number of rows in $Z$
    \For{$i = 1, 2, \ldots, m$} 
        \State $\vect{a} = Z_{[:,i]}\in\{0,1\}^{c}$, the $i^{\mbox{th}}$ column of $Z$
        \State $\vect{b} = Z_{[:, j\neq i]}\ones_{(m-1)\times 1}\in\mathbb{N}^{c \times 1}$, the row sum of $Z$ excluding the $i^{\mbox{th}}$ column
        \State $\vect{v}={\bf 0}_{c\times 1}$ 
        \For{$j=1,\ldots,c$}
            \State $v_j = \begin{cases} (2{a}_j-1)w_{\chi}({b}_j), \mbox{ if } {b}_j\le\chi-1 \mbox{ or } {b}_j\ge m-\chi \\ 0, \mbox{ otherwise} \end{cases}$
            \Statex \quad\quad\quad where $
		w_{\chi}({b}_j) = \begin{cases}
		\frac{{b}_j!(m-{b}_j-1)!}{m!}\cdot\frac{m}{2\chi}, \quad  \chi < \lceil m/2\rceil\\
		\frac{{b}_j!(m-{b}_j-1)!}{m!}, \quad\mbox{otherwise}.
		\end{cases}$
        \EndFor
        \State $\phi_i = \vect{v}^{\top}\ve$
    \EndFor
    \State $\phi_0 = f(\vq) - \sum_{i=1}^{m}\phi_i$
    \State {\bf output:} $\phi_0,\phi_1,\ldots,\phi_m$
\end{algorithmic}
\label{alg:hyperShap}
\end{algorithm}
\vskip -0.2in

\section{Preliminary Results}
\label{sec:experiments}

\subsection{Clinical Data}
The first phase of this project aims to improve the response assessment to systemic therapy of CRLM patients by applying advanced analytics to medical imaging and clinical data. All patient data used were collected as part of the multicenter randomized clinical trial CAIRO5~\cite{huiskens2015treatment}. This ongoing study aims to downsize tumor burden in the liver and make local treatment with curative intent feasible for initially unresectable colorectal liver metastases, in order to improve (disease-free) survival.  The data consisted of diagnostic imaging (CT images) before and after systemic therapy, evaluated by a nationwide expert-panel. The CT-images from 52 patients were used for segmentation of the liver and liver metastases by an expert radiologist. The expert segmentations were performed semi-automatically using the Philips\textsuperscript{\textregistered} IntelliSpace Portal software. A total of 1380 liver metastases were segmented, resulting in the 3-D organ contours of the liver and all metastases. From each tumor, each three-dimensional pixel (voxel) is available for analytics. 

\subsection{Model Training and Interpretation}

We built a deep-learning image segmentation model targeting labeled lesion regions on the CT images. For each new test CT image, we use the deep-learning model to predict lesion regions, and then calculate the Shapley values for the deep-learning model on the detected lesion regions in the image. Positive Shapley values indicate pixels that contribute positively towards the predicted probability of a lesion, while negatively Shapley values indicate pixels that contribute negatively towards the model's predicted probability of a lesion. By viewing the area that contributes towards the predicted probability of a lesion, a clinician can see what area of the CT image the deep-learning model thinks is indicative of a lesion in the liver.

\begin{figure}[ht!]
\centering
\includegraphics[width=.48\textwidth]{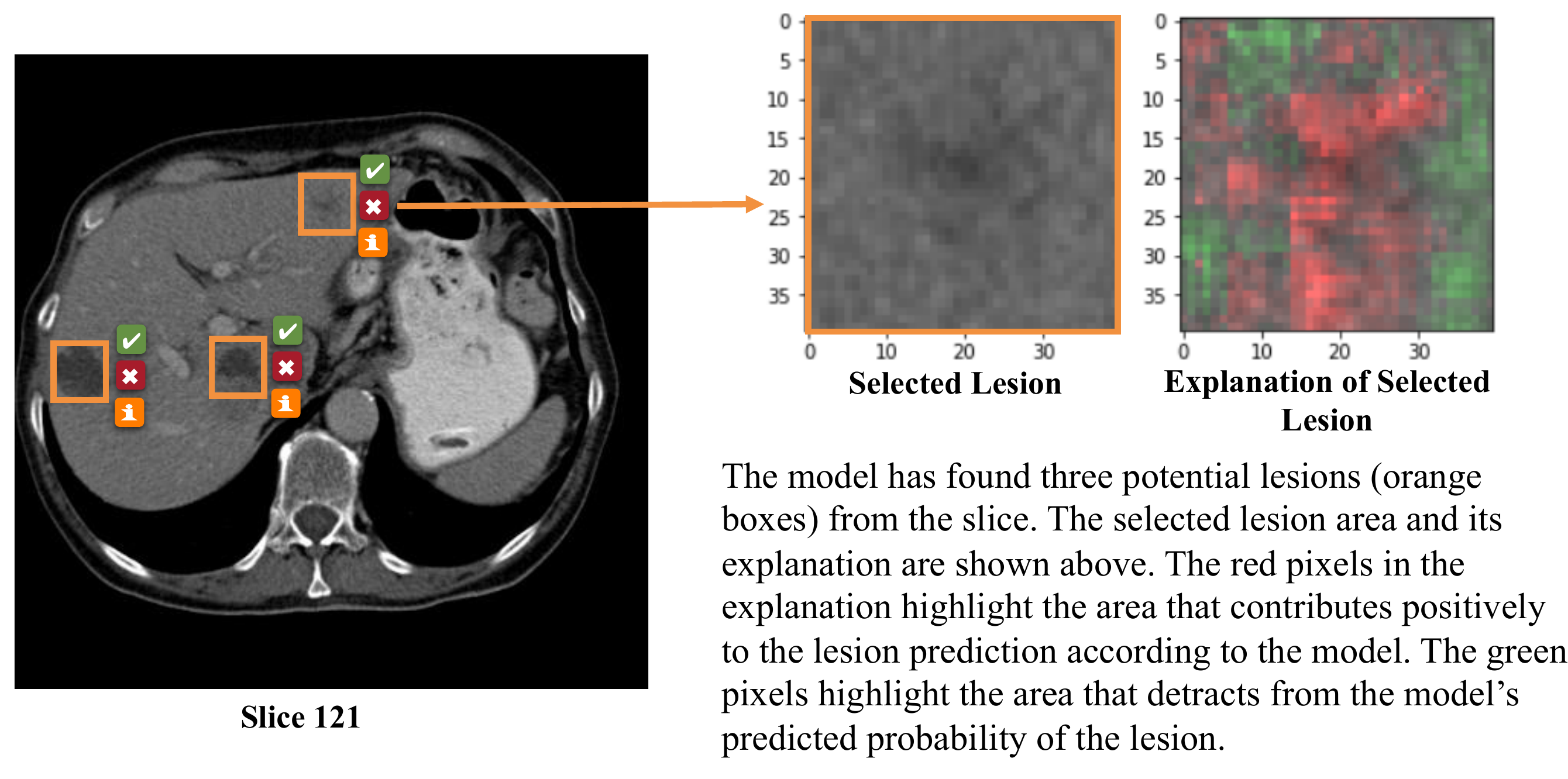}
\caption{A sample report. The clinician can confirm or reject each detected lesion by clicking the green or red button next to the bounding box. Clicking the orange button (labeled with letter ``i'') will show explanations for the corresponding image patch for review. The clinician can also label new areas as lesions and request explanations.}
\label{fig:report}
\vskip -0.2in
\end{figure}

In Figure~\ref{fig:report} we show a sample report generated using the proposed framework.

\section{Conclusion}
\label{sec:conclusion}
Current lesion detection and measurement systems are not
clinician-efficient, taking large amounts of clinicians' time. These systems also suffer from significant inter-clinician variability. The proposed system optimizes the use of clinicians' time by quickly identifying potential lesions and
providing interpretation as to why the model provided such a prediction. By automating a portion of the lesion detection task, our framework can reduce inter-clinician variability.

Our initial experiments have shown promise both in the ability to detect
lesions and the ability to explain the predictions of a lesion
detection model. Future work includes improving the detection model, improving the interaction system with clinicians, and
validating our AI-augmented framework through clinical trials.


\bibliographystyle{ACM-Reference-Format}
\bibliography{hyperShap}

\end{document}